\documentstyle[12pt]{article}
\begin{document}

\baselineskip=0.85truecm

\title{Conformal Cosmology and the Age of the Universe\footnote
{UCONN 95-08, December 1995}}

\author{\normalsize{Philip D. Mannheim} \\
\normalsize{Department of Physics,
University of Connecticut, Storrs, CT 06269} \\
\normalsize{mannheim@uconnvm.uconn.edu} \\}

\date{}

\maketitle

\begin{abstract}
We show that within the cosmology associated with conformal gravity the age of
the universe is given as $1/H_0$, to thus overcome the current cosmological age
crisis. We show that while the parameter $\Omega_{mat} = \rho_{mat} / \rho_C$
takes on all values between zero and infinity in conformal gravity, nonetheless
it is of order one (but not identically equal to one) for half a Hubble time to
thus naturally explain its current closeness to one without fine tuning. We
show
that the cosmological constant is naturally of order the energy density
$\rho_{mat}$ of ordinary matter again without fine tuning. We compare and
contrast conformal cosmology with that of the standard Friedmann cosmology.

\end{abstract}

For a theory which has long since been declared to be the true and correct
cosmological theory, the standard Friedmann model is currently beset by a
surprisingly large number of problems. The definitive new Hubble Space
Telescope
(Freedman et al 1994) determination of the value of the Hubble parameter $H_0$
causes the standard $\Omega_{mat}=1$, $\Omega_{vac}=0$ Friedmann model age
$t_0=2/3H_0$ for the universe to now be less than that of some of its
constituents. While the age prediction for the model can be increased by
allowing
for a non-zero cosmological constant vacuum energy contribution $\Omega_{vac}$
(see e.g. Krauss and Turner 1995 who actually argue for such a non-zero value
from a variety of cosmological considerations), unless this contribution is
constrained according to the inflationary universe requirement
$\Omega_{tot}=\Omega_{mat}+\Omega_{vac}=1$, the celebrated flatness problem
would then reappear. However, the fine tuning problem which would then be
required of $\Omega_{vac}$ to enforce this desired $\Omega_{tot}=1$ would then
be
no less than 60 or so orders of magnitude more severe than the flatness problem
tuning problem for which inflation was proposed in the first place (Guth 1981).
New cosmological data are thus forcing cosmologists to finally have to confront
the one problem that they had previously side stepped by setting
$\Omega_{vac}=0$ (not that any reason had been advanced for that choice
either).
Beyond these already quite severe issues, new abundance determinations and
computational analyses are calling into question (White et al 1993, Hata et al
1995, Copi, Schramm and Turner 1995) what had always been regarded as the
primary achievement of the standard model, namely big bang nucleosynthesis.

While it is of course much too early to draw any definitive conclusions
regarding
the ultimate status of the standard model, nonetheless the current situation
does
demand a critical reappraisal of its basic ingredients, with the most basic of
all being its reliance on the use of Newton-Einstein gravity in the first
place,
an issue which Mannheim and Kazanas have actually been challenging in a recent
series of papers simply by noting that there is currently no known principle
which would uniquely select out the Einstein-Hilbert action from amongst the
infinite class of all order covariant metric based theories of gravity that one
could in principle at least consider. Motivated by the fact (Mannheim 1990)
that the assumption of an underlying conformal symmetry (viz. invariance under
local conformal transformations of the form $g_{\mu \nu} (x) \rightarrow
\Omega (x) g_{\mu \nu} (x)$ and the consequently unique conformal invariant
gravitational action $I_W=-\alpha \int d^4x (-g)^{1/2} C_{\lambda\mu\nu\kappa}
C^{\lambda\mu\nu\kappa}$) actually strictly forbids the presence of any
fundamental cosmological constant (to thereby provide a symmetry based
framework
with which to address this longstanding problem), Mannheim and Kazanas then
embarked on a detailed analysis of the possible astrophysical implications of
the
conformal gravity theory. They solved for the exact exterior (Mannheim and
Kazanas 1989; see also Riegert 1984) and interior (Mannheim and Kazanas 1994)
metrics associated with a static, spherically symmetric source in the model,
demonstrated their consistency, and found that in conformal gravity all the
classic tests of General Relativity could still be met (even despite the
absence
of the Einstein-Hilbert action which is also forbidden by the conformal
invariance - to incidentally thereby demote the Planck length $L_{Pl}$ from
fundamental status and decouple it from quantum gravity fluctuations). Further,
it was shown that the theory actually departs from the
standard theory on galactic distance scales in a manner (Mannheim 1993a,
1995) which can provide for a resolution of the galactic rotation curve
problem without the need to introduce any dark matter, this dark matter of
course
being the primary and still totally elusive component of the standard
$\Omega_{mat}=1$ paradigm. Moreover, a first conformal cosmological model was
constructed (Mannheim 1992) and it was shown to naturally resolve the flatness
problem by necessarily possessing a $k \ll 0$ and thus far from flat topology.
Subsequently (Mannheim 1995) it was found that this very negative curvature
acts universally on the galaxies which make up the Hubble flow to completely
explain the departures of their rotation curves from the luminous Newton
expectation without any need for dark matter, while also automatically
enforcing
the universal Tully-Fisher relation. Moreover, an actual value for $k$, viz.
$k=-3.5 \times 10^{-60}$ cm$^{-2}$, was even extracted from the rotation curve
data. In the present paper we continue the study of Mannheim (1992) to
calculate
the age of the Universe in the model and to show that the current value of
$\Omega_{mat}$ is naturally of order one.

While we have already noted that the very fact of conformal invariance forces
us
to change the gravitational side of the gravitational equations of motion, with
the Bach tensor (the variation of the conformal action $I_W$  with respect to
the
metric) then replacing the familiar Einstein tensor, in a sense a possibly even
more far reaching aspect of conformal invariance is that it also forces us to
change the structure of the matter energy-momentum tensor side as well, thereby
forcing us to reconsider (Mannheim 1993b, Mannheim and Kazanas 1994)
the prevailing Newtonian 'billiard ball' perfect fluid view of gravitational
sources familiar in the standard applications of gravity to astrophysical
situations. Specifically, since conformal symmetry would require strictly
massless matter fields, it is necessary to introduce scalar (Higgs) fields
whose
non-vanishing vacuum expectation values would then spontaneously break the
conformal symmetry to give masses to the matter fields. Such scalar fields
would
then also carry energy and momentum which also couple to gravity (this energy
and momentum is simply ignored in the standard 'billiard ball' model of
sources,
even in fact when the Higgs fields are taken to be the conventional
Weinberg-Salam fields which are currently thought to generate particle masses);
and as we shall see, because of the underlying conformal structure, the
contributions of these scalar fields to the energy-momentum tensor prove to not
only be too significant to be ignored, but they also turn out to be constrained
in a way which enables us to address many current astrophysical puzzles.

To see what the constraints of conformal symmetry explicitly entail, consider
the typical case of fermionic matter fields Yukawa coupled to scale breaking
scalars. For them, the most general covariant, conformal invariant matter
action
$I_M$ takes the form
\begin{equation}
I_M=-\int d^4x(-g)^{1/2}[\hbar S^\mu S_\mu/2+\lambda S^4-
\hbar S^2R^\mu_{\phantom{\mu}\mu}/12+
i\hbar \bar{\psi}\gamma^{\mu}(x)(\partial_\mu+\Gamma_\mu(x))\psi -
hS\bar{\psi}\psi]
\label{Eq. (1)}
\end{equation}
where $\Gamma_\mu(x)$ is the fermion spin connection and $h$ and $\lambda$ are
dimensionless coupling constants. (For simplicity we only consider fermion
bilinears in $I_M$. In a more detailed model we would also need to consider
fields which transform as fermion quadrilinears as well with those fields being
responsible for scale breaking in the massless, high temperature, cosmological
radiation era.) When the scalar field $S(x)$ in $I_M$ acquires a non-vanishing
vacuum expectation value (which we are free to set equal to a spacetime
independent constant $S$ because of the conformal invariance), the fermion then
obeys the Dirac equation $i\hbar \gamma^{\mu}(x)[\partial_{\mu} +\Gamma_\mu(x)]
\psi = h S \psi$ and acquires a mass $hS$. Once $S$ is constant, we note that
this Dirac equation is the same as the one which is used for fermions with
mechanical masses (viz. 'billiard balls'), so that the geodesic motion for
massive fermions which follows from it is the standard one. For macroscopic
purposes we note that the incoherent averaging of the fermion kinetic energy
operator $i\hbar \bar{\psi}\gamma^{\mu}(x)(\partial_\mu+\Gamma_\mu(x))\psi$
over
all the occupied positive frequency modes of this Dirac equation leads us
(Mannheim 1992) to a standard kinematic perfect fluid of these fermions with
energy-momentum tensor $T^{\mu \nu}_{kin}=(\rho_{mat}+p_{mat})U^\mu U^\nu+
p_{mat}g^{\mu\nu}$; while the averaging of the total energy-momentum tensor
and of the scalar field equation of motion associated with Eq. (\ref{Eq. (1)})
lead us to
\begin{eqnarray}
T^{\mu\nu}=(\rho_{mat}+p_{mat})U^\mu U^\nu+p_{mat}g^{\mu\nu}
-\hbar S^2(R^{\mu\nu}-g^{\mu\nu}R^\alpha_{\phantom{\alpha}\alpha}/2)/6
-g^{\mu\nu}\lambda S^4
\nonumber \\
=(\rho_{mat}+p_{mat})U^\mu U^\nu+p_{mat}g^{\mu\nu}
-g^{\mu\nu}(3p_{mat}-\rho_{mat})/4
-\hbar S^2(R^{\mu\nu}-g^{\mu\nu}R^\alpha_{\phantom{\alpha}\alpha}/4)/6
\label{Eq. (2)}
\end{eqnarray}
and
\begin{equation}
\hbar S^2R^\mu_{\phantom{\mu}\mu}-24\lambda S^4+6(3p_{mat}-\rho_{mat})=0
\label{Eq. (3)}
\end{equation}
respectively. It is important to note that the $-g^{\mu\nu}(3p_{mat}-
\rho_{mat})/4$ term displayed in the second form of Eq. (\ref{Eq. (2)})
arises from the incoherent averaging of the Yukawa
$-g^{\mu\nu}hS\bar{\psi}\psi/4$ term and is needed to maintain the
tracelessness
of the full conformal $T^{\mu\nu}$, ($T^{\mu \nu}_{kin}$ itself of course is
not
traceless). Since the total $T^{\mu\nu}$ is also covariantly
conserved, we see from the first form of Eq. (\ref{Eq. (2)}) that
$T^{\mu \nu}_{kin}$ is conserved all on its own, with the sum of all the other
terms in the total $T^{\mu \nu}$ being independently conserved also. Thus all
the
standard features that arise from the covariant conservation of
$T^{\mu \nu}_{kin}$ (such as the dependence of the cosmological $\rho_{mat}$ on
$R(t)$) continue to hold in the conformal theory, with the motions of the
matter
fields being exactly the same as they would have been had the matter fields in
fact been billiard balls. However, since gravity couples to the entire
$T^{\mu \nu}$ and not merely to $T^{\mu \nu}_{kin}$, its behavior is radically
affected by the presence of all these additional non-Newtonian terms in the
full $T^{\mu \nu}$, and so we now explore their implications for cosmology.

For applications of conformal gravity to cosmology we note that in geometries
such as Robertson-Walker which are conformal to flat the conformal Bach tensor
($\delta I_W / \delta g_{\mu \nu}$) vanishes identically, so that the matter
fields are constrained to obey the equation of motion $T^{\mu \nu}=0$. Given
Eqs. (\ref{Eq. (2)}) and (\ref{Eq. (3)}), we thus see that since $T^{\mu
\nu}=0$
in conformal cosmology, the terms in $T^{\mu \nu}$ that depend on $S$ (which
collectively constitute a general cosmological term which includes both a
cosmological constant and a back reaction on the geometry) must between them
add
up to the energy density in ordinary matter, i.e. the magnitude of the
macroscopic $S$ is fixed by how many fermion states are occupied in
$\rho_{mat}$
in the first place. Thus we find that not only does the conformal theory
possess
no fundamental cosmological term, the one which is subsequently induced by the
symmetry breaking scalar field adjusts itself self-consistently via the back
reaction of the scalar field on the geometry to acquire a scale which is fixed
by
the energy density of the matter which got its mass from the selfsame scalar
field, so that the full cosmological term is thus neither smaller nor larger in
magnitude than the energy density of ordinary matter, to thus naturally fix
the positive frequency mode contribution to $\Omega_{vac}$
without fine tuning. Conformal cosmology thus naturally addresses
the issue of the magnitude of the cosmological term by using its symmetry
constraints. This situation should be contrasted with that of the standard
model,
a model which has no such constraints, and in which the self-consistent
back reaction on the geometry is not even considered - in fact the standard
model cosmological term is identified as $g^{\mu \nu} V_{min}(S)$ where
the minimum value $V_{min}(S)$ of the potential is simply transported from flat
space without regard to any of the other terms present in Eq. (\ref{Eq. (2)})
or to their mutual self-consistency. As we now see, the cosmological constant
problem should not in fact be viewed as a generic problem for cosmology, but
rather as a specific feature of the Einstein Equations, with the issue for the
standard theory being how to get rid of a term which has no reason not to be
there.

As regards the value of $\Omega_{mat}$ in the model, we note that in a
Robertson-Walker geometry the condition $T^{\mu \nu}=0$ reduces to
\begin{equation}
\dot{R}^2(t) +{2R^2(t)c\rho_{mat} \over \hbar S^2}=-kc^2-{2R^2(t)\lambda S^2c^2
\over \hbar}
\label{Eq. (4)}
\end{equation}
to yield a condition which only differs from the analogous standard model
equation in one regard, namely that the quantity $-\hbar S^2 /12$ has replaced
$c^3/16 \pi G$ in the second term on the left hand side. From the point of
view of the standard model, Eq. (\ref{Eq. (4)}) would have been obtained in
standard gravity if standard gravity were repulsive rather than attractive,
with the back reaction of the scalar field on
the geometry in conformal gravity thus acting like an induced effective
repulsive
rather than attractive gravitational term in the conformal case. Because of
this
crucial change in sign, the $\dot{R}^2 $ and $2R^2c\rho_{mat} /\hbar S^2$ terms
are required to add in Eq. (\ref{Eq. (4)}) rather than cancel so that the fine
tuning flatness problem present in the standard model (where these two huge
quantities have to cancel to an extraordinary degree of accuracy) is thus not
encountered in the conformal case (Mannheim 1992). Moreover, Eq. (\ref{Eq.
(4)})
can only have solutions at all if $k \ll 0$ (unless the coefficient
$\lambda$ is overwhelmingly negative, with it in fact generally even being
believed to be positive), and thus leads us to an automatically open and very
negatively curved Universe. (Essentially the only way the geometry can cancel
the positive energy density of ordinary matter and maintain $T^{\mu \nu}=0$ is
if the gravitational field itself contains the negative energy associated with
negative spatial curvature.) Now while we have shown that the conformal model
naturally avoids the flatness problem, nonetheless $\Omega_{mat}$ is still
quite close to one today, a fact which also must be natural in our model, and
it
is to this issue which we now turn.

It is most straightforward to discuss the general implications of Eq. (\ref{Eq.
(4)}) for $\Omega_{mat}$ in the simplified situation in which $\lambda$ is set
equal to zero. Since the matter era does not appear to possess solutions which
can readily be expressed in terms of elementary functions,
it is simpler to consider the radiation era. On setting $\rho_{mat} =A/R^4=
\sigma T^4$, Eq. (\ref{Eq. (4)}) is readily integrated to yield
\begin{equation}
R^2(t)=-kc^2t^2+R_{min}^2~~~,~~~
H(t)={1 \over t(1-R_{min}^2/kc^2t^2)}~~~,~~~
q(t)={R_{min}^2 \over kc^2t^2}
\label{Eq. (5)}
\end{equation}
so that $R(t)$ has a finite minimum radius $R_{min}$
$=(-2A/\hbar kS^2c)^{1/2}$ at $t=0$ (and thus a finite
maximum temperature $T_{max}$) with the cosmology thus being singularity free
(precisely because it induces a repulsive gravitational component so that
conformal gravity can protect itself from its own singularities - something of
course not the case in standard gravity). In terms of the conventionally
defined
and very convenient quantity $\Omega_{mat}(t)=\rho_{mat}/ \rho_C$, the
temperature at time $t$ then obeys
\begin{equation}
{T^2_{max} \over T^2(t)} =1-{kc^2t^2 \over
R_{min}^2}=1-{1 \over q(t)}=1+{4\pi L^2_{Pl}S^2 \over 3\Omega_{mat}(t)}
\label{Eq. (6)}
\end{equation}
\noindent
Thus for $\Omega_{mat}(t_0)$ currently of
order one the scale parameter $S$ must be at least as big as
$10^{10}L_{Pl}^{-1}$
if the early Universe is to have a maximum temperature of at least $10^{10}$
degrees. Analogously, the current value of the deceleration parameter must obey
$q(t_0)\leq 10^{-20}$. (To get larger phenomenological values for $q(t_0)$
would
require the reintroduction of the $\lambda$ term of Eq. (\ref{Eq. (4)}).) Then
since according to Eq. (\ref{Eq. (5)}) the Hubble and deceleration parameters
are
related as $H(t)(1-q(t))=1/t$ in the model, it follows that the age of the
Universe is given as $t_0=1/H(t_0)=1/H_0$, to be compared with $t_0=1/2H_0$ in
the standard model radiation era. In fact, as we show below, even in the matter
era the age remains $1/H_0$ in the conformal theory, and thus yields an age
which is currently phenomenologically viable.

Now according to Eq. (\ref{Eq. (6)}), $\Omega_{mat}(t)$ goes through all values
from infinity to zero during the lifetime of the Universe, and thus must pass
through one at some time, and as we have just shown there even exists a value
of
$S$ for which $\Omega_{mat}(t_0)$ is of order one today (though not identically
equal to one). Nonetheless, we still need to ask whether we are likely to be at
that value today since $\Omega_{mat}(t)$ could possibly be close to one only
for
a very short time, to then require some fine tuning to get it close to one in
the current epoch. To resolve this issue we note from Eqs. (\ref{Eq. (5)})
and (\ref{Eq. (6)}) that at time $t=t_0/2$ we obtain
\begin{equation}
{T^2(t_0/2) \over T^2(t_0)}={-kc^2t_0^2+R_{min}^2 \over
-kc^2t_0^2/4+R_{min}^2}={1-q(t_0) \over 1/4-q(t_0)}=4~~~,~~~
{\Omega_{mat}(t_0/2) \over \Omega_{mat}(t_0)}={t^2_0 \over t^2_0/4}=4
\label{Eq. (7)}
\end{equation}
so that both $T(t)$ and $\Omega_{mat}(t)$ take values close to their current
values for no less than half a Hubble time. Thus even though their early
Universe
values differ radically from their current values, the probability of finding
$T(t)$ and $\Omega_{mat}(t)$ in their current values at the current time is
still
very high. In this way the model explains why $\Omega_{mat}(t_0)$ can naturally
be of order one today despite its radically different values at very early
times.
In contrast, we recall that the flatness problem for the standard model stems
from the fact that given the closeness of $\Omega_{mat}(t_0)$ to one today, the
Friedmann evolution equations require $\Omega_{mat}(t)$ to be even closer to
one
at earlier times. Thus we see that the flatness problem is not in fact generic
to
cosmology, but rather it would appear to be a specific feature of the Einstein
Equations, and may thus even be a signal that the Einstein Equations might not
be
the appropriate ones for cosmology.

It is also of interest to see how the evolution equation of Eq. (\ref{Eq. (4)})
itself manages to avoid any fine tuning problem. In the solution of
Eq. (\ref{Eq. (5)}) the two terms on the left hand side of Eq. (\ref{Eq. (4)})
respectively take the form:
\begin{equation}
\dot{R}^2(t)={k^2c^4t^2 \over -kc^2t^2 +R_{min}^2}={-kc^2 \over (1-q(t))}
{}~~~,~~~{2Ac \over \hbar S^2 R^2(t)}={-kc^2R_{min}^2 \over -kc^2t^2+R_{min}^2}
={kc^2q(t) \over (1-q(t))}
\label{Eq. (8)}
\end{equation}
to thus give radically different time behaviors to these two terms even while
their sum remains constant (=$-c^2k$). Specifically, $\dot{R}^2$ begins	at zero
and slowly goes to $-c^2k$ at late times, while $2Ac/ \hbar S^2 R^2$ does the
reverse as it goes to zero from an initial value of $-c^2k$. Moreover,
evaluating them today then shows that both the terms have already attained
their late values, and that rather than being of the same order of magnitude
today, in fact the $\dot{R}^2(t_0)$ term is $10^{20}$ orders of magnitude
larger
than the $2Ac/\hbar S^2 R^2(t_0)$ term. This behavior differs radically from
that
found in the standard model (where the analogous two terms are both of the same
order of magnitude today) simply because the scale factor $S$ of the conformal
model is not of order $L_{Pl}^{-1}$ but rather a factor at least $10^{10}$
times
bigger. It is also of interest to ask at what time the two terms given in
Eq. (\ref{Eq. (8)}) were in fact of the same magnitude. From Eq. (\ref{Eq.
(8)})
we see that this would occur when $q(t)=-1$, i.e. at a time $t=t_0/10^{10}$, a
time at which $T(t)=T_{max}/\surd 2$ which is well in the early Universe; and
in
passing we note that in the conformal case the Universe initially cools very
slowly dropping in temperature by a factor of only $\surd 2$ in its first
$10^7$
sec.

It is also possible to extend the age estimate for the Universe to the matter
era where $\rho_{mat}=B/R^3$, a relation which fixes the magnitude of $S$ anew
in
accord with Eq. (\ref{Eq. (3)}). In this era the Universe is found to have a
minimum radius given by $R_{min}=-2B/kS^2\hbar c$ (we again set $\lambda=0$ for
simplicity), and an evolution given by
\begin{equation}
(-k)^{1/2}ct= R_{min}~log\left({R^{1/2}+(R-R_{min})^{1/2} \over R_{min}^{1/2}}
\right)+R^{1/2}(R-R_{min})^{1/2}
\label{Eq. (9)}
\end{equation}
and
\begin{equation}
{R(t) \over R_{min}}={T_{max} \over T(t)}= 1+{ 4\pi L^2_{Pl}S^2 \over
3\Omega_{mat}(t)} ={-kc^2 \over -kc^2-H^2(t)R^2(t)}
\label{Eq. (10)}
\end{equation}
\noindent
Thus again we find that $S \gg L_{Pl}^{-1}$ if $T_{max}$ is to be very big,
with
$R_0^2H_0^2$ then being extremely close to $-kc^2$ today. However, since $R_0$
is
very much greater than $R_{min}$, it follows from Eq. (\ref{Eq. (9)}) that
$-kc^2t_0^2 =R^2_0$ today, so that the age of the Universe is again given as
$t_0=1/H_0$ as required. Since the standard model matter era yields an age
$t_0=2/3H_0$, we see that conformal gravity yields an age which is 50\% bigger,
so that its age prediction for a Hubble parameter $h =0.75$ (defining
$H_0=100h$ km sec$^{-1}$ Mpc$^{-1}$) is the same as that of a standard model
with
the now excluded value of $h=0.5$ ; with the new HST value (Freedman et al
1994)
of $h=0.80 \pm 0.17$ actually yielding an age $1/H_0=12.2 \pm 2.6$ Gyr which
is compatible with the globular cluster age estimate of $16 \pm 3$ Gyr quoted
by
Krauss and Turner (1995).\footnote{While it is completely standard to compare
the age of the Universe with that of its constituents in the above way, it is
perhaps worth noting in passing that this procedure is only an approximate one,
and that its level of accuracy is only ascertainable by actually making a
general coordinate transformation between the time coordinate of the comoving
Robertson-Walker frame associated with the expansion of the Universe and that
of
each Schwarzschild coordinate rest frame system in which the age of each
constituent is measured.}

As we thus see, in both the radiation and matter eras the age of the Universe
is given as $t_0=1/H_0$, a result that could have been read off directly from
Eq. (\ref{Eq. (4)}) in the limit in which we drop the energy density
$\rho_{mat}$
altogether, viz. the pure curvature dominated limit in which
$\dot{R}^2(t)=-kc^2$
and $R(t)=(-k)^{1/2}ct$ (this also being a horizon free limit in which the
particle horizon $d(t)=R(t) \int_0^t dt /R(t)$ is infinite).\footnote{Moreover,
in the presence of matter straightforward calculation shows that for the form
of $R(t)$ given in Eq. (\ref{Eq. (5)}), the (dimensionless) ratio of the
horizon
size to the spatial radius of curvature $R_{curv}(t)$ ($=(6/R^{(3)})^{1/2}$
where $R^{(3)}$ is the modulus of the Ricci scalar of the spatial part of the
metric) is given as $d(t)/R_{curv}(t)=(-k)^{1/2}cd(t)/R(t)=log[T_{max}/T+
(T_{max}^2/T^2-1)^{1/2}]$ where $T_{max}=(4\pi /3\Omega_{mat}(t_0))^{1/2}T(t_0)
SL_{Pl}$ according to Eq. (\ref{Eq. (6)}). This ratio is thus much greater than
one at recombination to thus naturally solve the horizon problem in the
conformal
model. For comparison with the standard theory, we recall that when its
spatial curvature is negative, the standard theory yields
$d(t)/R_{curv}(t)=log[
T_{ref}/T+(T_{ref}^2/T^2+1)^{1/2}]$ ($T_{ref}=(1/\Omega_{mat}(t_0) -1)^{1/2}
T(t_0)$) a ratio which is much smaller than one at recombination. Comparing the
two expressions for the ratio we see that they only differ substantially at
recombination because the conformal inverse length scale $S$ is much greater
than $L_{Pl}^{-1}$, a feature which also enabled us to resolve the flatness
problem as we discussed above. As we thus now see, the origin of both the
flatness and horizon problems in the standard theory stems from the fact that
cosmological observables are apparently not naturally parameterized in terms of
the inverse length scale associated with Newton's constant, but rather by one
which is orders of magnitude bigger.} In fact this curvature
dominance drives the cosmology, a fact that could have been anticipated from
Eq. (\ref{Eq. (4)}), with this curvature dominance causing the Universe to
expand
far more slowly than in the standard case. While this curvature dominated
cosmology is thus seen to be able to address some outstanding puzzles of the
standard model, it is not itself yet completely free of problems, since this
same
slow expansion seems to be able to only produce substantial amounts of
primordial
helium and appears to have trouble generating other light elements
(Knox and Kosowsky 1993; Elizondo and Yepes, 1994). Whether this
is simply a property of using just the simple cosmology based on Eq. (\ref{Eq.
(1)}) and/or whether it could be resolved in more detailed dynamical conformal
models remains to be addressed. However, since the standard cosmology is also
having nucleosynthesis problems (and the standard cosmology has yet, despite
the prevailing view on dynamical generation of particle masses, to explain
exactly why it models the entire self-consistent mass generating $T^{\mu \nu}$
of Eq. (\ref{Eq. (2)}) purely by its mechanical 'billiard ball' kinematic
perfect
fluid $T^{\mu \nu}_{kin}$ piece),\footnote{To understand the exact nature of
this assumption, we note that from purely geometric considerations alone the
most general energy-momentum tensor in a Robertson-Walker geometry must take
the generic form $T^{\mu \nu}=(A(t)+B(t))U^{\mu}U^{\nu}+B(t)g^{\mu \nu}$, with
the coefficients $A(t)$ and $B(t)$ in general not necessarily being obliged to
be proportional to each other. Relations between $A(t)$ and
$B(t)$ only follow from specific dynamical assumptions. Moreover, if $A(t)$ and
$B(t)$ contain two or more separate components (this being the case for the
conformal $T^{\mu \nu}$ of Eq. (\ref{Eq. (2)})), then even if the separate
components are related via $A_1=w_1B_1$, $A_2=w_2B_2$, it does not follow
that $A_1+A_2$ is proportional to $B_1+B_2$. Further, even for the restricted
case of the kinematic $T^{\mu \nu}_{kin}$ where $A(t)$ and $B(t)$ are in fact
associated with a standard perfect fluid, it turns out that they are still not
in fact proportional to each other at all temperatures. Specifically, consider
an ideal $N$ particle classical gas of particles of mass $m$ in a volume $V$ at
a temperature $T$. For this system the Helmholtz free energy $A(V,T)$ is
given as exp$[-A(V,T)/NkT]=V\int d^3p$ exp$[-(p^2+m^2)^{1/2}/kT]$,
so that the pressure takes the simple form $P=-(\partial A/ \partial
V)_T=NkT/V$,
while the internal energy  $U=A-T(\partial A/ \partial T)_V$ evaluates in terms
of Bessel functions as $U=3NkT+Nm K_1(m/kT)/K_2(m/kT)$. In the two
limits $m/kT \rightarrow 0$, $m/kT \rightarrow \infty$ we then find that
$U \rightarrow 3NkT$, $U \rightarrow Nm+3NkT/2$. Thus only at these two extreme
temperature limits does it follow that the energy density and the pressure are
in fact proportional, with their relation in intermediate regimes such as the
transition region from the radiation to the matter era being far more
complicated. Since we thus see that the generic $A(t)$ and $B(t)$ are not in
fact
proportional to each other in general and at all temperatures, it would be
interesting to see to what degree the standard model nucleosynthesis
predictions are sensitive to more general choices for $A(t)$ and $B(t)$ than
conventionally considered.} more detailed study of this issue might prove
fruitful; and since the conformal theory does seem to be able to nicely address
so many other outstanding cosmological puzzles in such a straightforward manner
it would appear to merit further study.

The author would like to thank M. Turner for asking the right question. This
work
has been supported in part by the Department of Energy under grant No.
DE-FG02-92ER40716.00.


\begin{thebibliography}{99}

\bibitem{Copi1995} Copi, C. J., Schramm, D. N., and Turner, M. S. 1995, Phys.
Rev. Lett. 75, 3981.

\bibitem{Elizondo1994} Elizondo, D., and Yepes, G. 1994, ApJ, 428, 17.

\bibitem{Freedman1994} Freedman, W. L., Madore, B. F., Mould, J. R.,
Hill, R., Ferrarese, L., Kennicutt Jr., R. C., Saha, A., Stetson, P. B.,
Graham, J. A., Ford, H., Hoessel, J. G., Huchra, J., Hughes, S. M., and
Illingworth, G. D  1994, Nature, 371, 757.

\bibitem{Guth1981} Guth, A. H. 1981, Phys. Rev. D23, 347.

\bibitem{Hata1995} Hata, N., Scherrer, R. J., Steigman, G., Thomas, D., Walker,
T. P., Bludman, S., and Langacker, P. 1995, Phys. Rev. Lett. 75, 3977.

\bibitem{Knox1993} Knox, L., and Kosowsky, A. 1993, "Primordial Nucleosynthesis
in Conformal Weyl Gravity", Preprint Fermilab-Pub-93/322-A, November, 1993.

\bibitem{Krauss1995} Krauss, L. M., and Turner, M. S. 1995, "The Cosmological
Constant is Back", Preprint CWRU-P6-95, April, 1995.

\bibitem{Mannheim1990} Mannheim, P. D. 1990, GRG, 22, 289.

\bibitem{Mannheim1992} Mannheim, P. D. 1992, ApJ, 391, 429.

\bibitem{Mannheim1993a} Mannheim, P. D. 1993a, ApJ, 419, 150.

\bibitem{Mannheim1993b} Mannheim, P. D. 1993b, GRG, 25, 697.

\bibitem{Mannheim1995} Mannheim, P. D. 1995, "Cosmology and Galactic Rotation
Curves", Preprint UCONN 95-07, November, 1995.

\bibitem{MannheimandKazanas1989} Mannheim, P. D., and Kazanas, D. 1989, ApJ,
342, 635.

\bibitem{MannheimandKazanas1994} Mannheim, P. D., and Kazanas, D. 1994, GRG,
26, 337.

\bibitem{Riegert1984} Riegert, R. J. 1984, Phys. Rev. Lett. 53, 315.

\bibitem{White1993} White, S. D. M., Navarro, J. F., Evrard, A. E., and Frenk,
C. S.
1993, Nature, 366, 429.



\end{thebibliography}
\end{document}